\documentclass[12pt,preprint]{aastex}

\usepackage{graphicx,psfig,natbib}

\begin{document}

\title {Where Are The M Dwarf Disks Older Than 10 Million Years?}

\author{Peter Plavchan, M. Jura, \& S. J. Lipscy$^{\dagger}$} 
\affil{UCLA Department of Physics and Astronomy; plavchan@astro.ucla.edu}
\affil{$\dagger$ Now at Ball Aerospace}

\begin{abstract}
We present 11.7-micron observations of nine late-type dwarfs obtained at the Keck I 10-meter telescope in December 2002 and April 2003.  Our targets were selected for their youth or apparent IRAS 12-micron excess. For all nine sources, excess infrared emission is not detected.  We find that stellar wind drag can dominate the circumstellar grain removal and plausibly explain the dearth of M Dwarf systems older than 10 Myr with currently detected infrared excesses.  We predict M dwarfs possess fractional infrared excess on the order of $L_{IR}/L_{*}\sim 10^{-6}$ and this may be detectable with future efforts.
\end{abstract}
\keywords{stars: circumstellar matter -- stars: late-type -- stars: winds, outflows} 

\section{INTRODUCTION}

Until $\sim$20 years ago, our knowledge of planets and their formation had been limited to our own solar system, with planets around other stars being the subject of fiction and hypothesis.   With recent advances in instrumentation and detection capabilities, we are now confirming the existence of planetary systems around other stars.   In 1983, the InfraRed Astronomical Satellite (hereafter IRAS) offered the first evidence of dusty debris, or debris disks, orbiting other stars \citep[]{zuckerman01}.   The dust is heated by the incident stellar radiation and re-emits this radiation in the infrared, which we can then detect as an ``excess'' of infrared flux.  The first extra-solar planets were indirectly detected around a millisecond pulsar in 1993 by analysis of the pulsar timing \citep[]{wolszczan92}.  Finally, the first jovian planets around stars like our sun were discovered in 1995 through the characteristic ``wobble'' or radial velocity variations induced in the star \citep[]{mayor95}.  Studying the evolution of these systems and their characteristics -- both directly and indirectly -- allow us to begin to answer fundamental questions about the existence of life in the universe.

M Dwarfs are the lowest mass, size, luminosity and temperature main sequence stars which comprise $\sim\!\!70\%$ of the total number of stars in our galaxy \citep[]{mathioudakis93}.  If planetary systems exist around M Dwarfs, they could represent the most abundant planetary systems in the galaxy.  Thus, determining or constraining the abundance of exo-planetary systems around M Dwarfs is important.  

About $15\%$ of nearby main-sequence stars exhibit far-infrared excess characteristic of cold circumstellar dust, or debris disks analogous to our Kuiper belt \citep[]{habing01}.    However, most M-type stars do not currently have detected infrared excesses despite several targeted and blind surveys \citep[]{song02,weinberger04,zuckerman01,liu04,fajardo00}.  Reported 12$\mu$m, 20$\mu$m and 25$\mu$m IRAS and ISO (Infrared Space Observatory) excess emission candidates have been largely demonstrated as false-positives when followed-up by smaller aperture ground-based observations \citep[]{zuckerman01,song02,fajardo00,aumann91}.

We have performed a search for 12$\mu$m infrared excess around candidate M Dwarfs indicative of extrasolar zodiacal dust.  The presence of such warm debris disks,  analogous to the zodiacal dust in our own inner solar system, could be an indirect marker for the presence of parent bodies -- rocky planetesimals, asteroids and comets that can generate dust.  

We present selection criteria for our targets in Section 2, and we present our observations in Section 3.  In Section 4, we present an analysis of the observations using stellar atmosphere models to estimate photospheric contributions to the observed infrared flux.  We present a model for circumstellar grain removal for M dwarfs in Section 5 to explain our results and the dearth of M dwarfs older than 10 Myr with observed infrared excess.  In section 6, we discuss the implications of this model and we present our conclusions in Section 7.

\section{SAMPLE SELECTION}

We constructed our sample from two selection criteria -- youth and apparent 12$\mu$m excess.  Our target list is given in Table 1.

\subsection{Youth}

\citet[]{spangler01} have found that for solar-type stars, the infrared excess varies as $t_{*}^{-1.76}$ where $t_{*}$ is the age of the star.  They present a simple model for asteroidal destruction to explain this result.  Extrapolating from the infrared excess produced by our own zodiacal debris, if we can identify M type stars which are $\sim$0.1 the age of the Sun ($t_{*}\sim500$ Myr) or younger, any existing infrared excess might be detectable with ground-based facilities.  Similarly, Liu et al. (2004) estimated that the dust mass, ${M_{d}}$, varies as $t_{*}^{-0.5}$ to $t_{*}^{-1}$ from debris disk masses estimated from sub-mm data, with a simple unweighted fit giving  ${M_{d}} \propto t^{-0.7 \pm 0.2}$.   Finally, from the lunar impact record one can infer that the amount of debris and resulting infrared excess can be significantly higher for young solar analogs \citep[]{chyba91}.  See \citet[]{decin03} for a re-examination of the time-dependence of the IR excess amplitude, which leads to a different conclusion than \citet[]{spangler01}.

In order to identify young, nearby M-type stars, we have used the results for the ${\beta}$ Pic moving group from \citet[]{zuckerman01b} and the ``rapid'' rotators from the survey by \citet[]{gizis02} to identify three targets.  The most obvious candidate in the ${\beta}$ Pic moving group is GJ 3305 \citep[]{song02} since AU Mic (GJ 803) is also member of this moving group and an M-type dwarf with an imaged debris disk \citep[]{kalas04,liu04b}.  The estimated age of GJ 3305 is 12 Myr, or 3 ${\times}$ 10$^{-3}$ $t_{\odot}$.  From \citet[]{gizis02}, GJ 3304 and GJ 3136 are ``rapid rotator'' stars with $v\sin i$ = 30 km s$^{-1}$, so these stars are likely younger than the Hyades and therefore about 0.1 the age of the Sun \citep[]{terndrup00}.

\subsection{Apparent 12$\mu$m Excess}

With the release of the 2MASS All Sky Survey, we cross-correlated this catalog with the IRAS Point Source and Faint Source Catalogues, similar and analogous to the work of \citet[]{fajardo00} .  We used the volume complete 676-source M Dwarf sample of \citet[]{gizis02} and the 278-source dwarf K and M-type sample identified in \citet[]{mathioudakis93}.  There is moderate overlap between these two samples.   We selected targets with unusually red K-[12$\mu$m] colors. We eliminated sources exhibiting a false color excess attributable to IRAS beam confusion, such as binaries or relatively bright nearby companions within 1'.  Additionally, with the exception of HU Del (GJ 791.2), known binary sources were discarded.     Our most promising six targets were observed.  With the exception of GJ 585.1, a K5 dwarf, the remaining five targets are M-type. 

\section{OBSERVATIONS}

We present 11.7$\mu$m observations of nine late-type dwarfs obtained at the Keck I 10-meter telescope in December 2002 and April 2003.   Table 1 lists the target stars, their distance, 11.7 ${\mu}$m observed photometry, 12 ${\mu}$m IRAS photometry, and derived 3-$\sigma$ upper-limits on the infrared excess from synthetic spectral fitting.  Table 2 lists the observations and integration times.  Figures 1 and 2 shows spectral fits, observations and available literature photometry.

These three targets selected for their youth -- GJ 3136, GJ 3304 and GJ 3305 -- were observed at Keck at 11.7$\mu$m using the Long Wavelength Spectrograph (LWS) in imaging mode in December 2002 \citep[]{jones93}.  On-source integrations times for all three targets were two minutes.  Standard IDL routines were used to reduce the data, and we wrote IDL routines to flux-calibrate the data.

For the particular observations in December 2002, it was discovered later by Varoujan Gorjian \citep[private comm.]{gorjian03} and confirmed by Keck staff that a screw that adjusts a mirror in the LWS detector had come loose and fallen out.  This allowed some movement of the optics as a function of telescope orientation during the time of our observations.  However, we observed standards within 1' of our targets for GJ 3304 and GJ 3305, and all three targets and standards were observed at airmasses $\leq\!1.2$.  Nevertheless, this potentially adversely affected the calibrations used in our observations, especially for GJ 3136, introducing a systematic error.  Thus, we treated the variance in the calibration of our standards as systematic errors rather than random, and adjusted our resulting uncertainties in our observations accordingly.  The net result was to roughly double our final calibration uncertainties.

The six targets selected for their apparent IRAS 12$\mu$m excesses were observed at Keck using LWS at 11.7$\mu$m in April 2003, using the same reduction techniques described above.  On source integration times ranged from 2 to 5 minutes.  In the cases of G 203-047 and HU Del, target acquisition and identification was performed with NIRC (Keck Near Infra-Red Camera) K-band imaging \citep[]{matthews94}.  

\section{ANALYSIS}

Photospheric flux was estimated by fitting PHOENIX NextGen stellar atmosphere models \citep[]{hauschildt99} to available UBVRIJHK$_{s}$ photometry, taking into account effective bandpasses and photometric uncertainties in finding the minimum $\chi^{2}$ in temperature and normalization.  The JHK$_{s}$ photometry were obtained from 2MASS All-Sky Survey and converted to flux densities based on the calibration by \citet[]{cohen03}, outlined in the 2MASS Explanatory Supplement \citep[ch VI.4a]{cutri03}.  UBV and R-I photometry, when available, were obtained from the Gliese catalogue. R magnitudes, when available, were obtained from the USNO-A V2.0 catalogue queried through Vizier \citep[]{ochsenbein00,gliese79,gliese91,monet98}.  Correcting USNO-R to Landolt-R magnitudes did not affect our model SED fits.  The IRAS 12${\mu}$m flux densities were color-corrected assuming blackbody emission, following the procedures outlined in the IRAS Explanatory Supplement \citep[ch.VI.C.3]{beichman88}.    The adopted color correction factors were chosen depending on the effective temperature $T_{eff}$ of each star, with a typical value of $\sim\!\!1.41$.

PHOENIX Nextgen models were used to fit the observed photometry rather than a blackbody, since the stellar SEDs are very different from that of a blackbody \citep[]{song02,mullan89}.  We assumed $log\:g = 4.5$ and solar metallicity, as is typical for main-sequence stars in the solar neighborhood \citep[]{dantona97,drilling00}.  Six of our nine targets are previously identified as flare stars, which primarily affect the U and B magnitudes during flaring.  However, the exclusion of U and B band magnitudes in our SED fits did not alter the derived temperatures for our targets.  Uncertainties in the predicted model flux at 11.7$\mu$m are derived from the resulting variance of the spectral fit as a function of the temperature ($\pm 100K$) and effective normalization ($\pm 5\%$) uncertainties around the minimum reduced $\chi^{2}$ fit.    

We find that the spectral models and photometric data are self consistent with one another, and consistent with no detected infrared excess at $11.7{\mu}m$ for all nine targets.   Our derived effective temperatures are in agreement with published spectral types when available.  We obtain an overall reduced $\chi^{2}$ of the model fits to the data of 0.95, and on average the 3-$\sigma$ upper limits correspond to a ratio of $F_{d,\nu}/F_{*,\nu} \sim30\%$ at 11.7$\mu$m, excluding GJ 4068.  This ratio limit is typical for current ground-based mid-infrared observational capabilities; we are limited in our analysis by our photometric quality rather than model-fitting uncertainties.

\subsection{Comments on the discrepancies between our ground-based observations and IRAS flux measurements.}

IRAS did not detect GJ 3304, GJ 3305 and GJ 3136 in the Point Source and Faint Source Catalogues, but as one can see in Figure 1 our measurements at 11.7$\mu$m are consistent with the predicted photospheric emission.  For GJ 4068, GJ 507.1, HU Del and GJ 585.1, we note that the color-corrected IRAS 12$\mu$m measurements are inconsistent with our 11.7$\mu$m measurements with a significance of $\sim$5.5-, 3-, 2-, and 2.5-$\sigma$ respectively.  The IRAS 12$\mu$m measurements are also inconsistent with the predicted photospheric emission at the $\sim$9-, 2.5-, 2-, and 3-$\sigma$ respectively.  It is unlikely that these discrepancies are due to the presence of Si emission in the IRAS 12$\mu$m bandpass, which would be missed by the LWS 11.7$\mu$m bandpass \citep[]{metchev04,gaidos04}.  For HU Del and GJ 4068 we suspect that this inconsistency is due to beam confusion from nearby sources, located $\sim$1.0 and $\sim$2.1 arc-minutes away, respectively. For GJ 507.1 and GJ 585.1, we suggest that the inconsistencies are probably due to statistical fluctuations and the ``Malmquist-bias'' for stars near the detection threshold \citep[]{song02}.

For G 203-047 and GJ 729, we note that the color-corrected IRAS 12$\mu$m measurements and our 11.7$\mu$m measurements are both consistent with one another and the predicted model photospheric emission.    Although G 203-047 and GJ 729 possessed unusually red K-[12$\mu$m] colors in our initial sample, this was not indicative of a true excess with subsequent modeling of their cool photospheres.  For GJ 4068, the flux error is dominated by systematic error in our calibration rather than statistical uncertainties in the detection itself.  Thus, we report a detection rather than an upper-limit. For HU Del, we note that it is a spectroscopic binary, which may account for the model spectral fit appearing to be slightly inconsistent with the literature photometry.  Finally, for GJ 585.1, the only K dwarf in our sample, our 11.7$\mu$m measurement is above the predicted model photosphere at the 1.5$\sigma$ level, and may warrant further observations at longer wavelengths.  However, given that this is only a 1.5$\sigma$ excess, we do not believe it is likely to be real.

\section{MODELING OF CIRCUMSTELLAR PROCESSES}

In section 5.1, we review the current observational knowledge of M dwarf circumstellar environments.  In section 5.2, we contrast observations with a steady-state application of P-R grain removal for M Dwarfs.  In section 5.3 we present a model for stellar wind drag grain removal for M dwarfs.  In section 5.4, we discuss the current knowledge of M dwarf stellar winds.  In section 5.5, we then evaluate the relevance of stellar wind drag in M dwarf debris disks and we present a model for their evolution.  In section 5.6, we discuss other dust removal mechanisms.

\subsection{The Observational Dearth of M Dwarf Debris Disks}

Circumstellar disks are common around M Dwarfs younger than $\sim$5 Myr; the inner parts of these disks appear to dissipate within 5-10 Myr, and any remaining disks are relatively rare in systems older than 10 Myr.  AU Mic (GJ 803) and GJ 182 are the only two M-type dwarfs older than 10 Myrs with an infrared or sub-mm excess, confirmed with ground-based observations, that is indicative of a debris disk \citep[]{song02,liu04,zuckerman01,fajardo00}.   This lack of late-type dwarfs older than 10 Myrs with an infrared excess does not appear to be a selection effect due to IRAS detection limits but rather an age dependent phenomenon \citep[private comm.]{song02,song03}.  \citet[]{weinberger04} propose that the lack of warm infrared excesses associated with the inner disks of K and M Dwarfs in the TW Hya association implies rapid planet formation within 5-10 Myrs, which in turn depletes available parent bodies for dust replenishment.  For AU Mic, \citet[]{liu04} argue that the lack of warm (T $\sim$ 200K) dust in the inner region of the circumstellar disk suggests the presence of an unseen planetary companion, also indicative of rapid planet formation within $\sim\!\!10$ Myr.   

In support of this rapid planet formation theory, it is known that known that primordial material to potentially form planets is common around low-mass stars and brown dwarfs with ages less than a few million years \citep[]{liu03,klein03,pascucci03,lada00,beckwith90,osterloh95}.  Hen 3-600 (TWA 3, a multiple star system) and TWA 7 are M-type pre-main sequence members of the $\sim$5-10 Myr old TW Hydrae Association \citep[]{zuckerman01}.  Coku Tau 4, AA Tau and CD $-40^{\circ}8434$ are further examples of T Tauri type pre-main sequence M Dwarfs with observed infrared excesses \citep[]{metchev04,quillen04,hartmann95}.  However, recent simulations of terrestrial planet formation around solar type stars take on average 10-50 Myrs to accrete into the analogs of our solar system terrestrial planets.  This process includes large-scale impacts -- like the one that is theorized to have formed Earth's moon -- lasting on the time-scale of $\sim$100 Myrs \citep[]{agnor99} and thus providing parent bodies to generate extrasolar zodiacal clouds on these timescales.  The arguments of \citet[]{weinberger04,song02,liu04} lead us to conclude that there potentially exists a different debris disk evolution timescale for K and M Dwarfs, and herein we propose such a mechanism.  

\subsection{Grain Removal from Poynting-Robertson Drag}

The dust we observe in a debris disk is replenished by the grinding (destruction) of parent bodies \citep[]{zuckerman01}, and removed under the action of Poynting-Robertson and other forces.    Thus, the infrared excess can be related to the destruction rate of the parent bodies.  In a simple model, we derive from \citet[]{chen01} and \citet[]{jura04} a relation between L$_{IR}$ and the rate at which mass is being removed from parent bodies and converted into dust, ${\dot {M_{d}}}$:

\begin{equation}
{\dot M_{d}}\,c^{2}\;=\;C_{0}\,L_{IR}
\end{equation}
where $C_{0}$ is a numerical constant of order unity, $L_{IR} \approx (\nu L_{\nu})_{IR}$ is the luminosity of the dust infrared excess and $c$ is the speed of light.  \citet[]{chen01} used $C_{0}$ = 4, where $C_{0}$ depends upon the assumed initial dust distribution relative to the inner radius at which the dust sublimates.

We have used the steady-state assumption that ${\dot {M_{d}}} = M_{dust}/t_{removal}$ and Equations (4) and (5) in \citet[]{chen01} to arrive at Equation (1) above.  Note that in the derivation of \citet[]{chen01}, the dominant grain removal process is assumed to be Poynting-Robertson (P-R) drag ($t_{removal} = t_{PR}$).

For M-type dwarfs, the characteristic grain removal time-scales from Poynting-Roberston drag are significantly longer than for earlier type dwarfs due to the relatively lower luminosities in late-type dwarfs.    Thus, contrary the observational dearth of M-type debris disks, we might expect a large abundance of dust around these stars.

\subsection{Grain Removal from Corpuscular Stellar Wind Drag}

We propose that the corpuscular (proton) stellar wind drag, hereafter stellar wind drag, in late-type dwarfs serves as the dominant mechanism for grain removal in these stars, in contrast to the P-R drag cited for earlier-type debris disk evolution.  An analogous and more detailed derivation applied to red giants can be found in \citet[]{jura04}. 

The stellar wind drag force is caused primarily by the proton flux from the solar wind impacting dust grains and the resulting anisotropic recoil; the effect is analogous to the Poynting-Robertson drag \citep[]{burns79,gustafson94,holmes03}.   In our own solar system, the solar wind drag has been measured to be on the order of 30\% of the Poynting-Robertson drag force, varying in strength between 20\% and 43\% \citep[]{gustafson94}.  \citet[]{holmes03} approximates a value of 1/3.

To first order in $v_{orb}/c$ and $v_{orb}/v_{sw}$ -- assuming each of these quantities is $\ll 1$ as in \citet[]{gustafson94}, where $v_{orb}$ is the grain orbital velocity of a grain assumed to be in a circular orbit, and where $v_{sw}$ is the proton stellar wind velocity assumed to be entirely radial -- we can write the ratio of drag times $t_{pr}/t_{sw}$ in terms of the magnitude of force ratios:

\begin{equation}
\frac{t_{pr}}{t_{sw}} = \left|\frac{\vec{F_{sw}}\cdot\hat{\theta}}{\vec{F_{PR}}\cdot\hat{\theta}}\right| = {\frac{c}{v_{sw}}}\times \left|\frac{\vec{F_{sw}}\cdot\hat{r}}{\vec{F_{rad}}\cdot\hat{r}}\right|
\end{equation}
where $F_{sw}$, $F_{rad}$, and $F_{PR}$ correspond to the solar wind pressure, radiation pressure and Poynting-Robertson drag forces respectively, and $\hat{\theta}$ and $\hat{r}$ correspond to the azimuthal and radial unit vectors in spherical coordinates.  We note that while the radiative stellar wind pressure can be dominated by the radiation pressure, the drag force ratio is enhanced by a factor of c/v$_{sw}$ in favor of the corpuscular drag force.  We then write:

\begin{equation}
\frac{t_{pr}}{t_{sw}} = \frac{c}{v_{sw}}\times \frac{Q_{sw}}{Q_{pr}}\left(\frac{\dot M_{sw}v_{sw}}{L_{*}/c}\right) = \frac{Q_{sw}}{Q_{pr}}\frac{\dot M_{sw} c^{2}}{L_{*}}
\end{equation}
where properties such as grain size, density and distance from the star drop out from the above expression, and where $Q_{sw}/Q_{pr}$ is the ratio of coupling coefficients.  For our own Solar System, taking $\dot M_{sw} \sim 2\times10^{-14}$ M$_{\odot}$ yr$^{-1}$ and assuming $Q_{sw}/Q_{pr} = 1$, Equation (3) evaluates to 37\%.  

Equation (3) presents a first-order expression to evaluate the relative importance of the P-R and stelar wind drags; more detailed expressions for the drag forces can be found in \citep[]{gustafson94}.    We deduce from Equation (3) that stellar wind drag can dominate P-R drag due to the lower M-dwarf luminosities relative to the Sun.  We write a more general expression for the inferred parent body destruction rate in cgs units, $\dot M_{d}$, by combining Equation (1) and Equation (3):

\begin{equation}
\dot M_{d} =   \left(\frac{C_{0}\,L_{IR}}{c^{2}}\right) \times \left(1+\frac{\dot M_{sw}c^{2}}{L_{*}}\right) 
\end{equation}
where we have assumed for simplification and clarity above that $Q_{sw}/Q_{pr} = 1$ and we note again that L$_{IR}$ represents the luminosity of the infrared excess due to dust.  We can write the above equation in terms of the observed fractional infrared excess as:

\begin{equation}
\frac{L_{IR}}{L_{*}} = \frac{\dot M_{d} c^{2}}{ C_{0} (L_{*} + \dot M_{sw} c^{2})}  
\end{equation}

Equations (4) and (5) limit to the following expressions when stellar wind drag is the dominant grain removal mechanism:

\begin{equation}
\dot M_{d} = \frac{C_{0}  \dot M_{sw} L_{IR}}{L_{*}} 
\end{equation}

\begin{equation}
\frac{L_{IR}}{L_{*}} =  \frac{\dot M_{d} }{  C_{0} \dot M_{sw}} 
\end{equation}

Scaling the P-R drag timescale (see Equation (5) in \citet[]{chen01}) by $t_{sw}/t_{pr}$ from the right hand side of Equation (3), we derive the grain removal time-scale for stellar-wind drag:

\begin{equation}
t_{sw} = \frac{4\pi a \rho_{dust} D^{2}}{3 Q_{sw} \dot M_{sw}}
\end{equation}

where D is the distance of the grain from the star. 

\subsection{Late-Type Dwarf Stellar Winds}

Empirical and semi-empirical arguments have shown that late-type dwarf wind mass loss rates can exceed the solar wind mass loss rate of $\sim\!2\times10^{-14}$ M$_{\odot}$ yr$^{-1}$ by factors ranging from $\sim$10 \citep[]{mullan92,fleming95,wargelin01,wargelin02,wood01,wood02} to a proposed $\sim$10$^{4}$ in the case of the M dwarf flare-star YZ Cmi \citep[]{mullan92}.  This enhanced stellar wind mass-loss is expected due to the hotter coronae and increased magnetic activity in active late-type dwarfs.  Recent models suggest that the mass-loss rates from late-type dwarfs are a few times $10^{-12}$ M$_{\odot}$ yr$^{-1}$ \citep[]{wargelin01,lim96,vanden97}; this predicted mass loss rate is still a factor of $\sim$100 times the solar value.

The task of remotely observing such winds (directly or indirectly) around other solar-type dwarfs has been a substantial challenge.  The measurements of wind rates were made indirectly from ``astrospheric'' absorption high-resolution Hubble Space Telescope spectra of Ly$\alpha$ absorption lines -- neutral interstellar hydrogen heated by the presence of a wind and the resulting interaction with the local interstellar medium.  From observations of G and K dwarfs, \citet[]{wood02} derived the relationship that $\dot m_{sw} \propto F_{X}^{1.15\pm0.20}$ from correlating observed stellar wind mass-loss rates with X-ray activity, where F$_{X}$ is the X-ray flux at the surface of the star and $\dot m_{sw}$ is the mass-loss rate per unit surface area.     This relation implies a mass loss rate, $\dot M_{sw}$, $\sim$1000 times that of the solar wind for the particular case of AU Mic \citep[]{wood02,huensch99}.

With substantial uncertainty, \citet[]{wood02} derive $\dot M_{sw} \approx \dot M_{\odot} (t/t_{\odot})^{-2.00\pm0.52}$ from stellar rotational velocity evolution and correlation with F$_{X}$.  \citet[]{wood02} caution that this relation may not accurately extrapolate for stars younger than $\sim$300 Myr, but the mass-loss will generally decrease with time from some maximum value.  Solar wind mass-loss rates as high as 10$^{3}$ times its present value have been hypothesized to account for the required luminosity and corresponding initial solar mass \citep[]{sackmann03,wood02}, but such a large enhancement in the mass-loss rate may not extend to young M-type dwarfs \citep[]{lim96,vanden97}.

\subsubsection{Proxima Cen}

No one has detected a wind from the nearest M Dwarf, Proxima Cen, but ascribing an appropriate upper-limit is a matter of debate.  \citet[]{wood01} placed an upper limit on $\dot M_{sw}$ of 0.2 times the solar wind mass-loss rate, contrary to the arguments of \citet[]{wargelin01,lim96,vanden97}.  However, the assumptions of \citet[]{wood01} about the intrinsic stellar L$\alpha$ line profile are cited as controversial in \citet[]{wargelin02}. From X-ray observations of charge exchange between the wind of Proxima Cen and neutral ISM gas, \citet[]{wargelin02} instead place a 3-$\sigma$ upper-limit constraint of 14 times the solar wind mass-loss rate, $\dot M_{sw}$.   More accurate estimates and measurements of M dwarf wind mass-loss rates will resolve these differences and uncertainties.

\subsection{The Relevance of Stellar Wind Drag for M Dwarf Debris Disk Evolution}

It seems plausible that M Dwarf wind mass-loss rates can exceed the solar value by at least a factor of 10.  Consequently, scaling from Equation (3), we find that stellar wind drag dominates P-R drag in M dwarfs.  We can now estimate  $L_{IR}/L_{*}$ when stellar wind drag is the dominant grain removal mechanism.   From Equation (7), using  $\dot M_{sw} = \dot M_{\odot} (t/t_{\odot})^{-2.00\pm0.52}$ from \citet[]{wood02}, estimating $\dot M_{d} = 4\times 10^{6} \left(t_{*}/t_{\odot}\right)^{-1.76\pm0.2}$ g/s, and setting $C_{0}=4$, we calculate:

\begin{equation}
\frac{L_{IR}}{L_{*}}\sim 9\times10^{-7} \; \left(\frac{t_{*}}{t_{\odot}}\right)^{0.24\pm0.6}  
\end{equation}

Equation (9) predicts that the frequency of M dwarf debris disks older than $\sim$10 Myr is roughly independent of age.  Furthermore, the predicted infrared excess ratio is below typical observational limits \citep[]{metchev04,spangler01,liu04}.  For our estimate of $\dot M_{d}$ used above, we inferred the time-dependence from \citet[]{spangler01,liu04}, and we adopted a proportionality constant of $C_{0}L_{IR}/c^{2}=C_{0}L_{*}f_{d}/c^{2} = 4\times 10^6$g/s from Equation (1).  For this proportionality constant for $\dot M_{d}$, we again set $C_{0}=4$, $L_{*}=L_{\odot}$ and we inferred $f_{d} = 2.5 \times10^{-7}$ at $t=t_{\odot}$ from Figure 2 in \citet[]{spangler01}.  This estimate for $\dot M_{d}$ is consistent with the dust replenishment rate for the zodiacal cloud, $\dot M_{d} \sim\!3\times 10^{6}$ g/s \citep[]{fixsen02}.

\subsection{Other Dust Removal Mechanisms}

Stellar wind drag can dominate P-R drag in the evolution of M dwarf debris disks when $t_{pr}/t_{sw}>1$, which occurs when $\dot M_{sw}/ \dot M_{\odot}$ \raisebox{-4pt}{$\stackrel{>}{\sim}$} 3$\times L_{*}/L_{\odot}$.  We can now put this result  in context with the role of collisions and the role of grain blowout, both through radiation pressure and stellar wind pressure.   

Dust grain blowout due to radiation pressure is irrelevant for M dwarfs, due to the lower luminosity and longer peak wavelength of radiation ($L_{*}\sim10^{-1\;.. -3}L_{\odot}$, $\lambda_{peak}>1\mu m$).   We present a simplistic spherical grain model to explain this result.  Assuming the absorption coupling coefficient $Q_{a}=1$ and the average grain density $\rho_{dust} = 2.5$ g cm$^{-3}$, we calculate from Equation (2) in \citet[]{chen01} a radiative blowout grain size radius,  a$_{blowout} \sim 0.1\mu$m for a M0 dwarf, with $L_{*} = 0.1 L_{\odot}$, $M_{*} = 0.5 M_{\odot}$.  Because $2\pi a_{blowout} / \lambda_{peak} < 1$ for all M Dwarfs, our assumption that $Q_{a} = 1$ is invalid and Equation (2) in \citet[]{chen01} is not applicable.  

When $2\pi a / \lambda < 1$, we write from Equation 7.7 in \citet[]{spitzer}:
\begin{equation}
Q_{a}(a,\lambda)= -8\pi \; \mbox{Im}\left(\frac{n^2-1}{n^2+2}\right)\frac{a}{\lambda}
\end{equation}
where n is the complex index of refraction that is material-dependent.  Equation (10) evaluates to $Q_{a}(a,\lambda)\sim a/\lambda$ for various silicates at wavelengths of $\sim$1$\mu$m \citep[]{dorschner95,ossenkopf92}.  We integrate $Q_{a}(a,c/\nu)L_{\nu}$ over $\nu$ for a stellar blackbody spectrum and derive $Q_{a}L_{*}\sim L_{*}\; a/\lambda_{peak}$.  The ratio of the radiation pressure to the gravitational attraction is then independent of grain size:

\begin{equation}
\beta \equiv \frac{F_{rad}}{F_{grav}} \approx \frac{3L_{*}}{16\pi c G M_{*} \rho_{dust} \lambda_{peak}}
\end{equation}
where G is Newton's gravitational constant, and c is the speed of light.  Taking $\rho_{dust} = 2.5$ g cm$^{-3}$,$L_{*}=0.1L_{\odot}$, $M_{*}=0.5M_{\odot}$, and $\lambda_{peak} = 1 \mu m$, we calculate $\beta \sim 0.05$. We conclude that $\beta<<1$ for all M dwarfs -- radiation pressure is insufficient to overcome gravitational attraction in expelling orbiting grains of any size.  The grain morphology for these small grains can be important relative to the blackbody approximation, and a more detailed calculation of the effective cross section to evaluate the relevance of radiative blowout has been done by \citet[]{saija03}.  The authors similarly conclude that the radiative pressure for small amorphous grains orbiting a 2700K M dwarf is negligible.

\citet[]{krist05} suggest that stellar wind pressure may play a role in the dissipation of dust around AU Mic.  We compute the stellar wind blowout radius to be:

\begin{equation}
a_{blowout} = \frac{3Q_{sw} \dot M_{sw} v_{sw}}{16\pi G M_{*} \rho_{dust}}
\end{equation}
where $v_{sw}$ is the stellar wind velocity, and any grains smaller than $a_{blowout}$ will be expelled under the action of stellar wind pressure. This can be derived directly by taking the ratio of the stellar wind and gravitational forces and setting them equal to 1, or by multiplying Equation (2) in \citet[]{chen01} by $|\vec{F_{sw}}\cdot\hat{r} / \vec{F_{rad}}\cdot\hat{r}|$ (see Equations (2) and (3)).

For a 0.5M$_{\odot}$ M0 dwarf with a solar mass-loss rate and solar wind velocity of $\sim$400km/s \citep[]{feldman77}, assuming $Q_{sw}=1$ and $\rho_{dust}=2.5$g cm$^{-3}$, the corresponding blowout radius is insignificant at $2\times$10$^{-4}\mu m$.  However, for a 0.5M$_{\odot}$ M0 dwarf with a mass-loss rate of $\sim$10$^{3} \dot M_{\odot}$, the corresponding blowout radius would be 0.2$\mu$m, increasing to 0.9$\mu$m for a 0.1M$_{\odot}$ late-M dwarf with a mass-loss rate of $\sim$10$^{3} \dot M_{\odot}$.  Thus, the relevance of stellar wind blowout increases for later spectral types, but is only relevant for extremely young or active M Dwarfs with potentially unrealistic mass-loss rates.  The validity of our assumption that $Q_{sw}=1$ for these grain sizes is also uncertain.

\citet[]{dominik03} suggest that all the known debris disk systems are collision-dominated and not P-R drag dominated.  This relies on the assumption that dust grains can be ground down by collisions until they are blown out by radiation pressure on a time-scale short compared to a drag time-scale.   This assumption does not hold for M Dwarf debris disks, since dust grains are not blown out by radiation pressure.  While M dwarf debris disks will be collisionally processed on short time-scales, stellar wind drag or grain growth is likely to still be the dominant grain removal process.  This modification predicts an excess of small grains for M dwarf debris disks relative to those around earlier-type stars and hence a relatively blue disk color.

\section{DISCUSSION}

\subsection{Some Implications of Stellar Wind Drag for M Dwarfs}

The time-dependence and steady-state assumptions of Equation (9) may be oversimplified, and the normalization is similarly uncertain by at least a factor of two.  Nonetheless, stellar wind drag offers an explanation for the lack of M-type debris disks older than 10 Myrs identified in \citet[]{song02,fajardo99,mamajek04}.  For the late-type dwarfs with confirmed infrared excesses in \citet[]{metchev04}, the estimated disk masses derived from the \citet[]{chen01} model should be adjusted higher by the ratio of $t_{pr}/t_{sw}$.   For the late-type dwarfs in the TW Hydrae association, the removal of grains by the stellar wind drag offers an alternative explanation to rapid planet formation put forth by \citet[]{weinberger04}.  The circumstellar disks and forming planetesimals may still exist, but the infrared excess emission can be effectively ``surpressed'' by the stellar wind drag relative to the effects of the Poynting-Robertson drag.  When gas is still present in the disk, the dynamics are more complicated and our model does not apply.

\subsection{AU Mic: An Application of the Model}

AU Mic is the only M-dwarf with an imaged circumstellar disk and infrared excess older than 10 Myr.   We propose that stellar wind drag offers an alternative explanation for the lack of warm dust grains observed near AU Mic.  The observed substructure and turnover in the slope of the radial surface brightness profile are naturally explained by the presence of a planetesimal disk in the inner $\sim$35 AU \citet[]{metchev05,krist05,liu04b}.   This inner disk could be produced by the stochastic destruction of parent-bodies, the subsequent generation of dust, and the removal of the dust by stellar wind drag within a time short compared to the $\sim$12 Myr age of AU Mic. If circumstellar gas is present in the AU Mic disk, the physical interpretation becomes more complicated and our model does not apply.

We can estimate grain removal time-scales from Equation (8).  We assume a grain size of a = 0.5$\mu$m  \citep[]{metchev05}, a grain density of $\rho_{s} = 2.5$ g cm$^{-3}$, and $Q_{sw} = Q_{pr} \sim 1$.  Since these assumptions dictate the grain opacity, which is unknown to within an order of magnitude, our estimates are similarly uncertain.   For AU Mic, we could take $\dot M_{sw}\sim$10$^{3} \dot M_{\odot}$ from the X-ray luminosity \citep[]{wood02,huensch99}.  However, this is a factor of 5-10 higher than the maximum $\dot M_{sw}$ inferred from modeling M dwarf winds \citep[]{wargelin01,lim96,vanden97}.  Thus, our estimate for $\dot M_{sw}$ is also uncertain by an order of magnitude, and we instead adopt $\dot M_{sw} = 10^2 \dot M_{\odot}$ for this calculation.  If $\dot M_{sw} = 10^{2} \dot M_{\odot}$, then stellar wind drag removes grains smaller than 0.5$\mu$m in the inner 35AU of the AU Mic disk within $\sim$$4\times10^4$yr.    Similarly, grains smaller than 0.5$\mu$m will be removed out to 250AU in $\sim$2Myr.  We deduce that at least the inner portion of the AU Mic debris disk must have been replenished by collisions of parent bodies.  Neglecting stellar wind drag for comparison and assuming $L_{*}=0.13L_{\odot}$ \cite[]{metchev05}, P-R drag would remove grains smaller than 0.5$\mu$m out to $\sim$35 AU in $\sim$8 Myr.   

If $\dot M_{sw} = 10^{3} \dot M_{\odot}$, we calculate a corresponding stellar wind blowout radius of 0.2$\mu$m from Equation (12), where we assume $Q_{sw}=1$, $\rho_{dust} = 2.5$ g cm$^{-3}$, $M_{*}=0.5M_{\odot}$, and $v_{sw}=v_{sw,\odot}=400$km/s.  This is consistent with the minimum grain radius of 0.3$\mu$m derived observationally by \citet[]{metchev05} at a distance of 50-60AU.   This implies there will be a relatively large number of grains less than $1\mu$m in size in the AU Mic disk, since such grains are too large to be blown out by stellar wind pressure and radiation pressure is negligible.   This is consistent with the observed overall blue color of the AU Mic disk relative to the colors of disks around earlier-type stars \citep[]{metchev05,krist05}.

Equation (9) predicts that $L_{IR}/L_{*}\sim 2 \times 10^{-7}$ for AU Mic, although we caution against applying Equation (9) to stars as young as AU Mic.  \citet[]{liu04} measure $L_{IR}/L_{*}\sim 6 \times 10^{-4}$, a factor of $3\times10^{3}$ higher, from infrared and sub-mm observations.  This implies that the AU Mic disk is not in a steady-state; leftover primordial material or a recent major planetesimal collision contributing to the observed dust excess could be considered ``transient'' effects. 

\section{CONCLUSIONS}

We do not observe any excess 11.7$\mu$m emission for nine late-type dwarfs, selected for their youth or apparent IRAS 12$\mu$m excess.  We find that stellar wind drag can dominate the Poynting-Robertson effect in grain removal from late-type dwarf debris disks, and this offers an explanation for the dearth of known M Dwarf systems older than 10 Myrs with infrared excesses.  We predict that M Dwarf debris disks older than 10 Myr will have a roughly equal age distribution, with $L_{IR}/L_{*}\sim10^{-6}$.  Future efforts, such as the Spitzer Space Telescope, may be successful in detecting these systems.

\subsection{Acknowledgements}

The authors wish to recognize and acknowledge the very significant cultural role and reverence that the summit of Mauna Kea has always had within the indigenous Hawaiian community.  We are most fortunate to have the opportunity to conduct observations from this mountain.  

This publication makes use of data products from the Two Micron All Sky Survey, which is a joint project of the University of Massachusetts and the Infrared Processing and Analysis Center/California Institute of Technology, funded by the National Aeronautics and Space Administration and the National Science Foundation.

This research has made use of the SIMBAD database, operated at CDS, Strasbourg, France.  This research has made use of the VizieR catalogue access tool, CDS, Strasbourg, France.
 
Thanks to Michael Liu, Ben Zuckerman, Inseok Song, Paul Kalas, Alycia Weinberger, Christine Chen, Stanimir Metchev and Thayne Currie for their insightful conversations and comments.  This work is supported by NASA.

\newpage
TABLES \\

Table 1.  Extrasolar Zodiacal Dust\\
\\
\begin{tabular}{||l|l|l|l|l|l|r||} \hline\hline
\emph{Name} & \emph{$T_{eff}^{(1)}$}  & \emph{D$^{(2)}\!\!$} &  \emph{$F_{11.7\mu m}\!\!$ obs.$\!\!$} & \emph{$F_{IRAS\:12\mu m}\!\!^{(3)}$$\!\!$} & \emph{$F_{11.7\mu m\!\!}$ mod.$\!\!$} & \emph{$F_{d,\nu}/F_{*,\nu}^{(4)}\!\!$}  \\
 & (K) & (pc) & (mJy) & (mJy)& (mJy) & \\ \hline

GJ 3136 & 3200 & 14.3 $\pm$ 2.9 & 43 $\pm$ 8 & --& 48 $\pm$ 2&$<$ 0.56\\
GJ 3305 & 3700 & 29.8 $\pm$ 0.3$^{(5)}$ & 106 $\pm$ 19& --& 103 $\pm$ 2&$<$ 0.54\\
GJ 3304 & 3100  & 10.0 $\pm$ 1.7 & 50 $\pm$ 9& --& 61 $\pm$ 2&$<$ 0.54\\
GJ 507.1 & 3600 & 19.8 $\pm$ 4.3 & 98 $\pm$ 3& 166 $\pm$ 23& 109 $\pm$ 3&$<$ 0.09\\
GJ 585.1 & 4200$^{(6)}$  & 25.6 $\pm$ 1.1 & 62 $\pm$ 5& 106 $\pm$ 18& 53 $\pm$ 2&$<$ 0.24\\
G 203-047 & 3200 & 7.3 $\pm$ 0.5 & 122 $\pm$ 5& 135 $\pm$ 15& 125 $\pm$ 4&$<$ 0.12\\
GJ 729 & 3100 & 2.97 $\pm$ 0.02 & 373 $\pm$ 8& 392 $\pm$ 74& 382 $\pm$ 14&$<$ 0.06\\
GJ 4068 & 3100 & 18.2 $\pm$ 4.3 & 22 $\pm$ 12$^{(7)}$& 102 $\pm$ 8& 30 $\pm$ 1&$<$ 1.64\\
HU Del & 3000 & 8.9 $\pm$ 0.2 & 55 $\pm$ 3& 89 $\pm$ 16& 59 $\pm$ 3&$<$ 0.16\\ \hline \hline
\end{tabular}\\
$^{1}$ From reduced $\chi^{2}$ fit of PHOENIX NextGen model spectra to nearest 100 K.\\
$^{2}$ Inferred from parallax obtained through SIMBAD and Vizier unless otherwise noted.\\
$^{3}$ Color-corrected as described in section 2.3, taken from the IRAS Faint Source Catalogue and Point Source Catalogue where available.\\
$^{4}$ 3-$\sigma$ upper limits inferred from 11.7$\mu$m observations as a function of frequency, $\nu$. \\
$^{5}$  \citet{zuckerman01b}; Inferred from SIMBAD parallax for primary companion, 51 Eri.\\
$^{6}$ Model spectra fitted to $\pm 200 K$.  The best fit temperature is consistent with this source's previously identified spectral type, K5.\\
$^{7}$ The error quoted for this particular flux measurement is dominated by systematic error in our calibration rather than statistical uncertainties in the detection itself.  Thus we report a detection rather than an upper-limit.

\newpage
Table 2.  Journal of 11.7$\mu$m Observations\\
\\
\begin{tabular}{||l|l|r|l||} \hline\hline
\emph{} & \emph{Date Observed}  & \emph{Integration} \\
\emph{Name} & \emph{(HST, night of)}  & \emph{Time (sec)} \\ \hline
GJ 3136 & 12/20/2002 & 120\\
GJ 3305 & 12/20/2002 & 120\\ 
GJ 3304 & 12/20/2002  & 120\\
GJ 507.1 & 4/23/2003 & 240\\
GJ 585.1 & 4/23/2003 & 300\\
G 203-047 & 4/23/2003 & 300\\
GJ 729 & 4/23/2003 & 144\\
GJ 4068 & 4/23/2003 & 120\\
HU Del & 4/23/2003 & 120\\ \hline \hline
\end{tabular}\\
\newpage
\begin{figure}
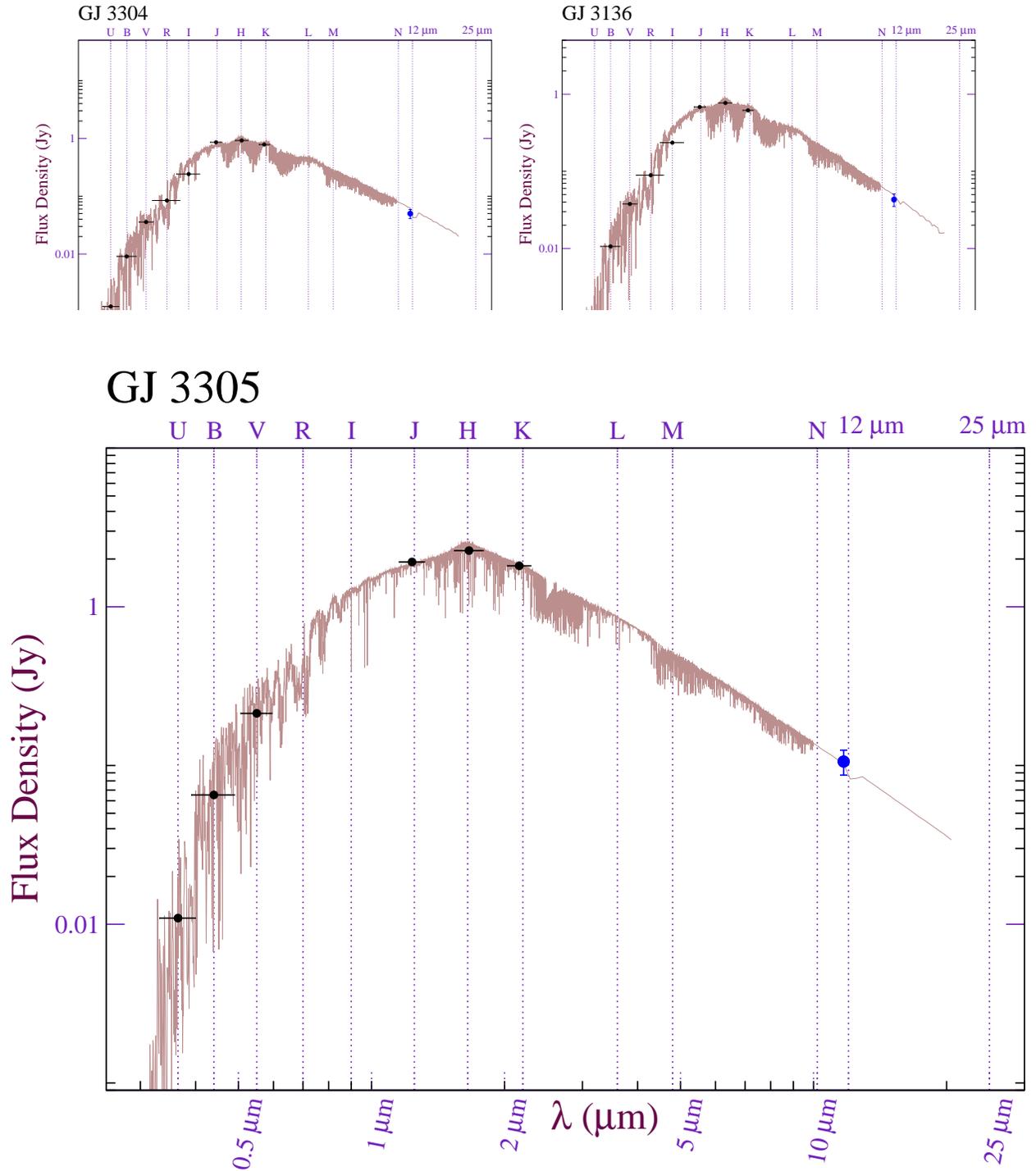

\epsscale{1}
\plottwo{f1a}{f1b}\\
\plotone{f1c}
\caption {Model spectral fits, photometric data and observations, December 2002.  Available UBVRIJHK$_{s}$ photometry and bandpasses are displayed in black.  11.7$\mu$m observations shown with error bars in blue.  Model spectra are displayed in brown.}
\end{figure}
\begin{figure}
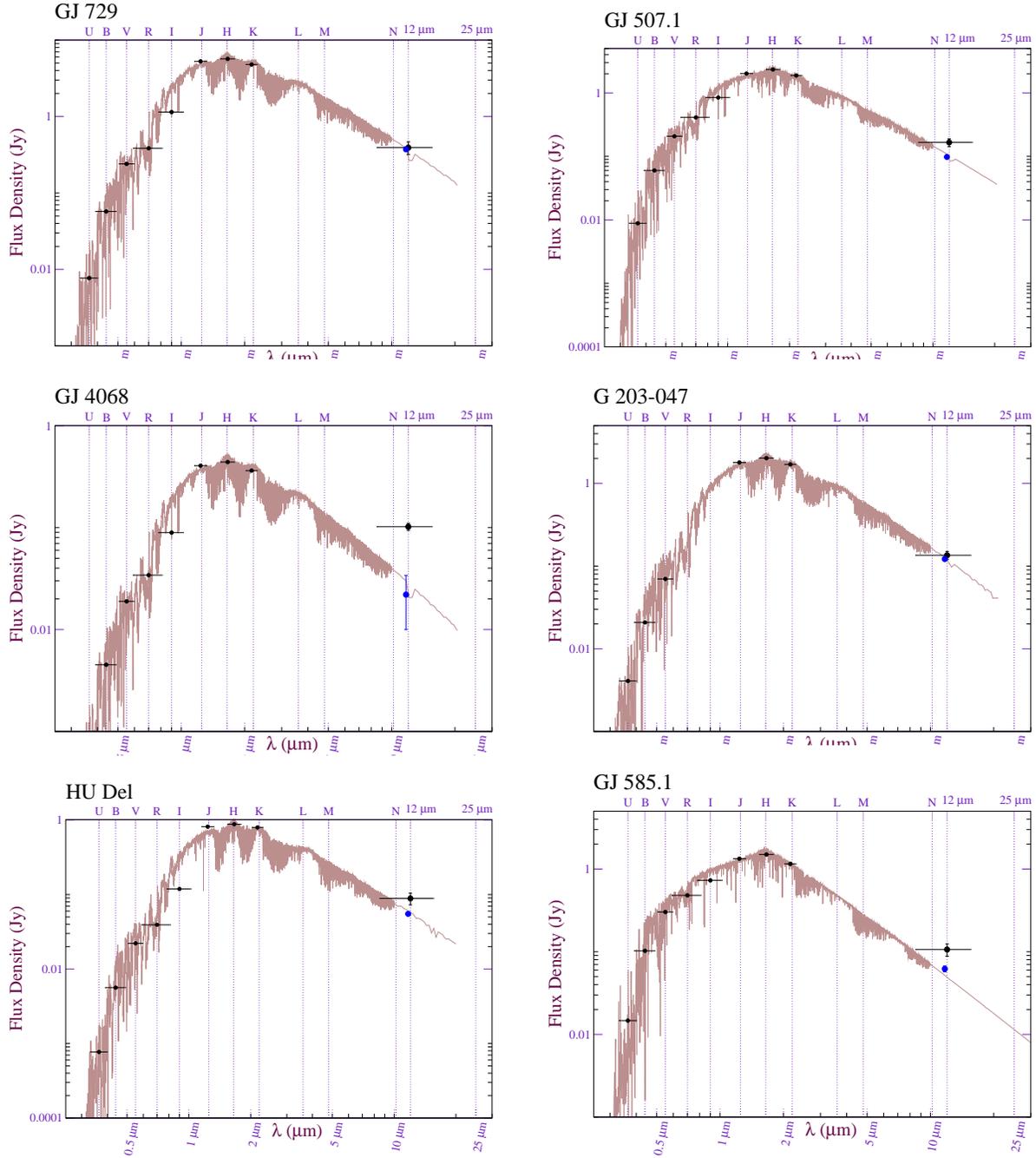

\epsscale{1}
\plottwo{f2a}{f2b}\\
\plottwo{f2c}{f2d}\\
\plottwo{f2e}{f2f}\\
\caption {Model spectral fits, photometric data and observations, April 2003.  Available UBVRIJHK$_{s}$ photometry and bandpasses are displayed in black.  12$\mu$m IRAS measurements, color-corrected, are also displayed in black with effective bandpass and errors shown.  11.7$\mu$m observations shown with error bars in blue.  Model spectra are displayed in brown.  See text for discussion.}
\end{figure}
\end{document}